\documentclass[a4paper,12pt]{article}
\pdfoutput=1
\usepackage{feynmp-auto,expdlist}
\usepackage{amsmath, amsfonts}
\usepackage{graphicx,arydshln}
\usepackage{enumerate}
\usepackage{hyperref}
\usepackage{latexsym}
\usepackage{dsfont}
\usepackage{hepnicenames}
\usepackage{enumerate}
\usepackage{soul}
\usepackage[normalem]{ulem}
\usepackage{comment}

\newcommand{\footnoten}[1]{}

\usepackage[font={small},textfont={it}]{caption} 

\usepackage{units}

\usepackage{mathrsfs,graphicx,rotating,amsmath,amsfonts,mathtools,booktabs,amssymb,wasysym}
\usepackage{hyperref}\usepackage{slashed}
\usepackage[nosort]{cite}
\usepackage[table,xcdraw,dvipsnames]{xcolor}
\usepackage{bm}
\usepackage{multirow,multicol}
\hypersetup{colorlinks,bookmarksopen,bookmarksnumbered,
	linkcolor=blus,pdfstartview=FitH,urlcolor=rossos,citecolor=verde}
\allowdisplaybreaks

\renewcommand{\[}{\left[}

\newcommand{\GF}{G_{\rm F}}

\def\Lag{\mathscr{L}}

\newcommand{\mio}[1]{}

\newcommand{\med}[1]{\langle #1\rangle}

\def\bpm{\begin{pmatrix}}
	\def\epm{\end{pmatrix}}

\usepackage{mathrsfs}

\newcommand{\fig}[1]{~\ref{fig:#1}}

\allowdisplaybreaks
\usepackage{multicol}
\usepackage{color}
\definecolor{rosso}{cmyk}{0,1,1,0.4}
\definecolor{rossos}{cmyk}{0,1,1,0.55}
\definecolor{rossoc}{cmyk}{0,1,1,0.2}
\definecolor{blu}{cmyk}{1,1,0,0.3}
\definecolor{blus}{cmyk}{1,1,0,0.6}
\definecolor{bluc}{cmyk}{1,1,0,0.1}
\definecolor{verde}{cmyk}{0.92,0,0.59,0.25}
\definecolor{verdec}{cmyk}{0.92,0,0.59,0.15}
\definecolor{verdes}{cmyk}{0.92,0,0.59,0.4}

\newcommand{\eq}[1]{~{\rm (\ref{eq:#1})}}

\newcommand{\s}{\,{\rm s}}

\newcommand{\MeV}{\,{\rm MeV}}
\newcommand{\GeV}{\,{\rm GeV}}
\newcommand{\TeV}{\,{\rm TeV}}
\newcommand{\cm}{\,{\rm cm}}

\def\circa#1{\,\raise.3ex\hbox{$#1$\kern-.75em\lower1ex\hbox{$\sim$}}\,}

\newcommand{\beq}{\begin{equation}}
\newcommand{\eeq}{\end{equation}}

\newcommand{\bea}{\begin{eqnarray}}
\newcommand{\eea}{\end{eqnarray}}
\newcommand{\be}{\begin{equation}}
\newcommand{\ee}{\end{equation}}
\font\tenrsfs=rsfs10 at 12pt
\font\sevenrsfs=rsfs7
\font\fiversfs=rsfs5
\newfam\rsfsfam
\textfont\rsfsfam=\tenrsfs
\scriptfont\rsfsfam=\sevenrsfs
\scriptscriptfont\rsfsfam=\fiversfs

\newsavebox\MBox

\renewenvironment{thebibliography}[1]
{\begin{multicols}{2}[\section*{\refname}]%
		\@mkboth{\MakeUppercase\refname}{\MakeUppercase\refname}%
		\list{\@biblabel{\@arabic\c@enumiv}}%
		{\settowidth\labelwidth{\@biblabel{#1}}%
			\leftmargin\labelwidth
			\advance\leftmargin\labelsep
			\@openbib@code
			\usecounter{enumiv}%
			\let\p@enumiv\@empty
			\renewcommand\theenumiv{\@arabic\c@enumiv}}%
		\sloppy
		\clubpenalty4000
		\@clubpenalty \clubpenalty
		\widowpenalty4000%
		\sfcode`\.\@m}
	{\def\@noitemerr
		{\@latex@warning{Empty `thebibliography' environment}}%
		\endlist\end{multicols}}

\newcommand{\eV}{\,{\rm eV}}

\def\circa#1{\,\raise.3ex\hbox{$#1$\kern-.75em\lower1ex\hbox{$\sim$}}\,}
\makeatletter

\font\ital=cmu10

\def\hhref#1{\href{http://arxiv.org/abs/#1}{arXiv:#1}}
\usepackage{xstring}
\newcommand{\hhrefq}[1]{\IfSubStr{#1}{:}{\href{http://inspirehep.net/search?ln=en&ln=en&p=#1&of=hb&action_search=Search&sf=&so=d&rm=&rg=25&sc=0}{InSpire:#1}}{\hhref{#1}}}

\def\art{\@ifnextchar[{\eart}{\oart}}
\def\eart[#1]#2#3#4#5#6{{\rm #2}, {\em #3 \bf #4} {\rm (#6) #5} ({\em #1})}
\def\article{\@ifnextchar[{\earticle}{\oarticle}}
\def\oarticle#1#2#3#4#5#6{{\rm #1}, {\ital ``#6''}, {\rm #2 #3 (#5) #4}}
\def\earticle[#1]#2#3#4#5#6#7{{\rm #2}, {\ital ``#7''}, {\rm #3 #4 (#6) #5}  [\hhrefq{#1}]}
\def\hepart[#1]#2{{\rm #2, \sl#1}}
\def\heparticle[#1]#2#3{#2, {\ital ``#3''} [\hhrefq{#1}]}
\newcommand{\doi}[1]{\href{http://dx.doi.org/#1}{[link]}}

\newcommand{\hhrefqq}[1]{\IfBeginWith{#1}{10.}{\href{https://doi.org/#1}{doi:#1}}{\hhrefq{#1}}}
\def\earticle[#1]#2#3#4#5#6#7{{\rm #2}, {\ital ``#7''}, {\rm #3 #4 (#6) #5}  [\hhrefqq{#1}]}

\renewenvironment{thebibliography}[1]
{\begin{multicols}{2}[\section*{\refname}]%
		\@mkboth{\MakeUppercase\refname}{\MakeUppercase\refname}%
		\list{\@biblabel{\@arabic\c@enumiv}}%
		{\settowidth\labelwidth{\@biblabel{#1}}%
			\leftmargin\labelwidth
			\advance\leftmargin\labelsep
			\@openbib@code
			\usecounter{enumiv}%
			\let\p@enumiv\@empty
			\renewcommand\theenumiv{\@arabic\c@enumiv}}%
		\sloppy
		\clubpenalty4000
		\@clubpenalty \clubpenalty
		\widowpenalty4000%
		\sfcode`\.\@m}
	{\def\@noitemerr
		{\@latex@warning{Empty `thebibliography' environment}}%
		\endlist\end{multicols}}

\newcounter{alphaequation}[equation]
\def\thealphaequation{\theequation\hbox to
	0.6em{\hfil\alph{alphaequation}\hfil}}
\def\eqnsystem#1{
	\def\@eqnnum{{\rm (\thealphaequation)}}
	\def\@@eqncr{\let\@tempa\relax \ifcase\@eqcnt \def\@tempa{& & &} \or
		\def\@tempa{& &}\or \def\@tempa{&}\fi\@tempa
		\if@eqnsw\@eqnnum\refstepcounter{alphaequation}\fi
		\global\@eqnswtrue\global\@eqcnt=0\cr}
	\refstepcounter{equation} \let\@currentlabel\theequation \def\@tempb{#1}
	\ifx\@tempb\empty\else\label{#1}\fi
	\refstepcounter{alphaequation}
	\let\@currentlabel\thealphaequation
	\global\@eqnswtrue\global\@eqcnt=0 \tabskip\@centering\let\\=\@eqncr
	$$\halign to \displaywidth\bgroup \@eqnsel\hskip\@centering
	$\displaystyle\tabskip\z@{##}$&\global\@eqcnt\@ne
	\hskip2\arraycolsep\hfil${##}$\hfil& \global\@eqcnt\tw@\hskip2\arraycolsep
	$\displaystyle\tabskip\z@{##}$\hfil
	\tabskip\@centering&\llap{##}\tabskip\z@\cr}
\def\endeqnsystem{\@@eqncr\egroup$$\global\@ignoretrue} \makeatother

\definecolor{Gray}{gray}{0.95}

\def\bal#1\eal{\begin{align}#1\end{align}}

\begin{document}

\thispagestyle{empty}

\begin{center}  
{\huge\bf\color{rossos} Dark Matter interpretation of\\[1ex]  the neutron decay anomaly}
{\huge\bf\color{rossos}}

\vspace{1cm}
{\bf\large Alessandro Strumia}  \\[6mm]
{\it Dipartimento di Fisica, Universit\`a di Pisa, Italia}\\[1mm]

\vspace{1cm}
{\large\bf Abstract}\begin{quote}\large
We add to the Standard Model a new fermion $\chi$ with minimal baryon number 1/3.
Neutron decay $n \to \chi\chi\chi$ into non-relativistic $\chi$ can account for the
neutron decay anomaly, compatibly with bounds from neutron stars.
$\chi$ can be Dark Matter, and its cosmological abundance
can be generated by freeze-in dominated at $T \sim m_n$.
The associated processes $n \to \chi\chi\chi \gamma$,
hydrogen decay ${\rm H}\to \chi\chi \chi\nu(\gamma)$ and
DM-induced neutron disappearance $\bar\chi n \to \chi \chi (\gamma)$ 
have rates below experimental bounds and can be of interest for future experiments.
\end{quote}
\end{center}

\setcounter{page}{1}
\tableofcontents

\section{Introduction}
The neutron life-time has been measured with two different experimental techniques,
which consistently give different results:
\begin{itemize}
\item The `bottle method' measures the total neutron decay width, with the result  
$\Gamma_n^{\rm tot} = 1/(878.3\pm 0.3)\s$, 
by storing ultra-cold neutrons in a magnetic bottle and counting their remaining
number after some time~\cite{nucl-ex/0408009,Pichlmaier:2010zz,Steyerl:2012zz,1412.7434,Arzumanov:2015tea,1707.01817,1712.05663,2106.10375}.
 
\item The `beam method' measures the $n\to p e\bar\nu_e$ $\beta$-decay rate,
with the result $\Gamma_n^\beta = 1/(888\pm 2)\s$, by 
counting the protons produced from a beam of cold neutrons~\cite{Byrne:1996zz,nucl-ex/0411041,1309.2623}.

\end{itemize}
These measurements contradict at about  $4.6\sigma$ level
the Standard Model (SM) that predicts $\Gamma_n^{\rm tot}=\Gamma_n^\beta$  up to sub-leading channels that can be neglected.
If confirmed, they indicate new physics.
It has been proposed that \beq\label{eq:DG}
\Delta\Gamma=\Gamma_n^{\rm tot} - \Gamma_n^\beta \approx \frac{1.2~10^{-5}}{\sec} (1\pm 0.21)\eeq
could be due to an extra decay channel of the neutron that does not produce protons.
The  new decay mode needs to be nearly invisible with a
branching ratio around $1\%$.
Various authors proposed a $n\to\chi \gamma$ decay into a new neutral fermion $\chi$ with mass $M$ slightly below the neutron mass so that $E_\gamma \approx m_n - M$ is small~\cite{1801.01124,2007.13931} (see also~\cite{Zurabtalk} and~\cite{1004.2981,2008.06061}).
However, this interpretation has the following problems:


\begin{enumerate}
\item First, if $M < m_p + m_e$ 
the new particle $\chi$ does not promptly decay back to SM charged particles and can be Dark Matter; 
but in this range $E_\gamma > m_n -m_p -m_e$ is large enough that
the extra decay is visible enough that it has been tested and disfavoured~\cite{1802.01595,1509.01223,2007.13931}.
An alternative that can explain both the neutron anomaly and Dark Matter (DM) would be welcome.

\item Second, $n\leftrightarrow\chi\gamma$ would lead to $\chi$ thermalization inside neutron stars,
as they are older than a Myr, while the neutron decay anomaly needs the much short time-scale of eq.\eq{DG}.
Thermalized $\chi$ soften the equation of state $\wp(\rho)$ of neutron stars.
Free $\chi$ add more energy density $\rho$ than pressure $\wp$ 
reducing the 
maximal mass of neutron stars below $0.7 M_{\rm sun}$,
like in the 1939 Oppenheimer and Volkoff's calculation that assumed free neutrons,
neglecting their nuclear self-repulsion~\cite{Oppenheimer:1939ne}.

\item Third, the precise SM prediction of the neutron decay rate, possibly
$\Gamma_n^{\rm SM} = 1/(878.7 \pm 0.6)\s$~\cite{1802.01804,Berezhiani:2018eds},  
agrees with the bottle experiments
disfavouring new physics contributions that increase the neutron decay rate. 
This might indicate that the neutron decay
anomaly could just be due to some under-estimated systematic uncertainty of `beam' experiments,
or that it needs alternative interpretations where the neutron oscillates (with a resonant enhancement thanks to magnetic fields in the apparata)
into a composite twin of the neutron~\cite{Berezhiani:2018eds,Berezhiani:2020zck,Berezhiani:2000gw}.\footnote{Among these models, those where the dark sector that is a mirror copy of the SM
also explain the near coincidence in mass needed to fit the anomaly, 
and were studed earlier because of their intrinsic theoretical interest~\cite{Lee:1956qn,
Kobzarev:1966qya,Blinnikov:1983gh,Kolb:1985bf,Foot:1991py,Hodges:1993yb,hep-ph/9511221,
hep-ph/9602326,1303.1727,1401.3965}.}
However, the quoted precise SM prediction for $\Gamma_n^{\rm SM} $ employs  determinations of $V_{ud}$ and  measurements of the neutron axial coupling constant $g_A$
that are potentially problematic at the claimed level of precision.
In particular, the pre-2002 measurements of $g_A$ gave instead a  $\Gamma_n^{\rm SM} $ value 
close to the beam experiments~\cite{1802.01804}. We thereby proceed ignoring this possible third problem.
\end{enumerate}
A proposed solution to the second problem is adding some new light mediator such that 
$\chi$ undergoes repulsive interactions
(with itself or with neutrons) stronger than the QCD repulsion among neutrons: this would make
energetically favourable to avoid producing $\chi$ in thermal equilibrium inside a neutron star, stiffening the equation of state~\cite{hep-ph/0507031,1802.08244,1802.08282,1802.08427,1811.06546}.
However this requires QCD-size  scattering cross sections that, if $\chi$ is DM, 
risk conflicting with the bullet-cluster bound on DM interactions.
Again, an alternative that can explain both the neutron anomaly and Dark Matter would be welcome.

\smallskip

In section~\ref{models} we discuss how more general models where new physics opens 
different neutron decay channels
affect neutron stars depending on the conserved baryon number $B_\chi$ assigned to $\chi$.
In section~\ref{NS} we show that  the minimal choice $B_\chi=1/3$ leads to
$n \to \chi\chi\chi$ decays into free DM $\chi$ that give a milder
modification of neutron stars, compatible with observations.
In section~\ref{TH} we elaborate on the theory behind and compute related processes.
In section~\ref{DMOmega} we show that $\chi$ can be DM, with abundance possibly
generated via freeze-in.
In section~\ref{DMsig} we show that $\chi$ DM has unusual signals compatible with current bounds.
Conclusions are presented in section~\ref{concl}.

\begin{table}[t]
$$\begin{array}{ccc|ccc}
\multicolumn{3}{c|}{\hbox{DM quantum numbers}} & \multicolumn{3}{c}{\hbox{DM interactions}}\\
 B & \phantom{-}L &  \hbox{spin} & \hbox{dimension} & \hbox{with quarks} & \hbox{with hadrons}  \\ \hline 
\rowcolor[rgb]{0.97,0.97,0.93}
  1 & \phantom{-}0 & 1/2 & 6 & \chi udd & \chi n \\ \rowcolor[rgb]{0.97,0.97,0.93}
 1/3 & \phantom{-}0 & 1/2 & 9 & \chi\chi\chi udd & \chi\chi\chi n \\ \rowcolor[rgb]{0.97,0.97,0.93}
 2/3 & \phantom{-}0 & 0 & 9 & \phi^3 (udd)^2 &\phi^3 n^2 \\ \rowcolor[rgb]{0.97,0.97,0.93}
  2 & \phantom{-}0 & 0 & 7 &\phi (udd)^2 & \phi nn \\  \rowcolor[rgb]{0.95,0.97,0.97} 
 0 & \phantom{-}1 & 1/2 & 4,6 & \chi L H, \chi \ell f \bar f & \chi  \ell \pi, \chi \ell  p\bar{n}\\  \rowcolor[rgb]{0.95,0.97,0.97}  0 & \phantom{-}2 &0 & 6, 8 & \phi (L H)^2,\phi  \ell \ell X q\bar q & \phi \nu \nu ,\phi \ell \ell \pi \pi   \\  \rowcolor[rgb]{0.99,0.97,0.97}
1 & \phantom{-}1 & 0 & 7 &  \phi L QQQ, \phi \ell uud & \phi n \nu, \phi p \ell\\ \rowcolor[rgb]{0.99,0.97,0.97}
1 & -1 & 0 & 8 & \phi \bar\ell X qqq & \phi n \bar \nu, \phi \Delta^- \bar\ell , \phi n \pi^- \bar\ell\\ \rowcolor[rgb]{0.99,0.97,0.97}
1 & \phantom{-}2 & 1/2 & 9 & \chi \ell \nu qqq & \chi n  \nu \nu,\chi p \ell \nu\\
 \end{array}$$
\caption{\label{tab:BLmodels}
Possible models of a new DM particle that carries baryon and lepton number $B,L$.
The left columns lists possible $B,L$ assignments, and the consequent minimal DM spin
(a scalar $\phi$ or a fermion $\chi$).
The right columns show representative examples of lowest-dimension effective operators
that couple DM to the SM conserving $B$ and $L$: here $f$ denotes a generic SM fermion,
either a generic quark $q$, or a charged lepton $\ell$ or a neutrino $\nu$.
$X$ denotes a derivative or a Higgs doublet $H$ (both have dimension 1).
$L $ is the SM lepton doublet that contains a neutrino, so that $L H$ contains $\nu v$.}
\end{table}

\section{Overview of possible models}\label{models}
To motivate the model we will propose, we first
classify  models where SM particles couple to one new particle,
by describing how baryon number $B$ and lepton number $L$ can be assigned to the new particle
such that they are conserved, in order to satisfy the strong phenomenological bounds
on $B, L$ violations. 

The $L$ and $B$ charges of the new particle determine its spin 
needed to couple to SM particles via operators with lowest dimension.
Since in the SM only fermions carry lepton and baryon number,
the new particle must have charge $(-1)^{3B+L}$
under the $\mathbb{Z}_2$ symmetry that flips signs of all fermions and
leaves bosons unchanged.
Table~\ref{tab:BLmodels} lists the main possibilities, focusing
on the minimal spin: either a fermion $\chi$ with spin 1/2, or a boson $\phi$ with spin 0
(or, equivalently, two fermions $\chi\chi$).
See~\cite{0806.0876,1410.4193,2005.00008,2007.07899,2007.08125} for lists of SM operators, partially relevant for our study.

\medskip

The apparently minimal choice (upper row of table~\ref{tab:BLmodels})
arises when a fermion
$\chi$ carries baryon number $B_\chi=1$ and vanishing lepton number $L_\chi=0$.
This is the model of~\cite{1801.01124,2007.13931}
where the neutron decays as $n\to\chi\gamma$. 
Models that involve a boson $\phi$ not protected by  Fermi repulsion 
worsen the neutron star problem.
Models where a fermion $\chi$ carries both lepton and baryon number 
do not avoid the neutron star problem, as neutrinos  freely escape from neutron stars.
Independently of the specific particle-physics interactions, the $\chi$ chemical potential $\mu_\chi$
in thermal equilibrium in a neutron star
is fixed in terms of the chemical potential of
conserved charges (baryon number and vanishing electric charge) as
\beq \label{eq:muchi}
\mu_\chi = B_\chi \mu_n.\eeq
%
Eq.\eq{muchi} shows that all models affect neutron stars in a qualitatively  similar way.
Eq.\eq{muchi} also shows how important quantitative differences can arise:
reducing $\mu_\chi$ allows to substantially reduce the impact on neutron stars.
This is achieved noticing that a more minimal choice of quantum numbers exists,
given that baryon number can be fractional, like the electric charge.
In the more minimal model $\chi$ carries  baryon number $B_\chi=1/3$ and $L_\chi=0$
(second row of table~\ref{tab:BLmodels}).
So neutrons decay as $n \to \chi\chi\chi$ if the new particle is light enough, $M < m_n/3$.
This neutron decay can have a small enough impact on neutron stars
for the same reason why ordinary neutron decay $n \to p e\bar\nu_e$ has a small impact:
the neutron decays to particles light enough that their Fermi repulsion is big enough,
without the need of introducing extra repulsive interactions.

\begin{figure}[t]
\begin{center}
\includegraphics[width=\textwidth]{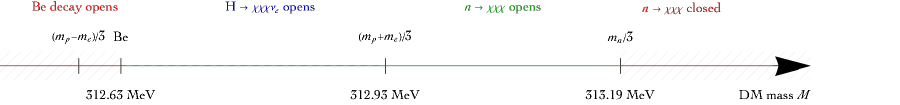} \end{center}
\caption{\label{fig:kinematics3chi} Kinematical thresholds for key decays
of a fermion $\chi$ with with baryon number $B=1/3$.}
\end{figure}


We thereby focus on this model. 
The DM mass $M$ is strongly restricted as illustrated in fig.\fig{kinematics3chi},
where we plot the following key kinematical thresholds:
\begin{itemize}

\item One needs $M < m_n/3 \approx 313.19\MeV$ so that $n \to \chi\chi\chi$ is kinematically open.

\item One needs $M > (m_n -E_{\rm Be})/3 \approx 312.63 \MeV $ 
so that nuclear decays into $\chi\chi\chi$ are kinematically closed,
where $E_{\rm Be}=1.664\MeV$ since the strongest bound comes from $^8$Be~\cite{1801.01124}.

\item
Proton stability gives the weaker bound $M> (m_p-m_e)/3 =312.59\MeV$.

\item Hydrogen decay ${\rm H} \to \chi\chi\chi\nu_e$ is kinematically open for
$M < (m_p + m_e)/3\approx 312.93\MeV$.
This part of the allowed parameter space leads to the possible signals discussed later.
\end{itemize}
The neutron decay anomaly can be explained in the mass range highlighted in
green in  fig.\fig{kinematics3chi},
while it cannot be explained in the red shaded range.

\section{Bounds from neutron stars}\label{NS}
Neutron stars are described by the
Tolman-Oppenheimer-Volkoff (TOV) equations~\cite{Oppenheimer:1939ne,astro-ph/0605724}
\beq 
\frac{d\wp}{dr} = - \frac{G}{r^2} \frac{({\cal M}+4\pi r^3 \wp)(\rho+\wp)}{1-2G{\cal M} /r},\qquad \frac{d{\cal M}}{dr}=4\pi r^2 \,\rho
\eeq
for the pressure $\wp(r)$ at radius $r$ and for the mass ${\cal M}(r)$ inside radius $r$.
The TOV equations describe spherical hydrostatic equilibrium in general relativity
and can be solved for any equation of state that gives the density $\rho$ in terms of $\wp$,
by starting with an arbitrary pressure at the center at $r=0$ where ${\cal M}(0)=0$, 
and by evolving outwards until finding the neutron star radius $r=R$ at which $\rho(R)=0$.
Repeating this procedure predicts the relation between the radius $R$
and the total mass ${\cal M}$, 
in particular giving a maximal mass ${\cal M}$ above which neutron stars become unstable and
collapse into black holes.
In the SM this maximal mass is around two solar masses,
compatibly with observations of neutron stars around this mass.

\medskip

Adding new stable particles $\chi$ to neutrons and to the other sub-dominant SM particles, 
the total energy density and pressure become
 $\rho=\rho_n + \rho_\chi$ and $\wp = \wp_n + \wp_\chi$.
The equation of state with the two components in equilibrium is computed as follows.
We assume that the $\chi$ particles are free fermions with mass $M$ and $g=2$ degrees of freedom, 
so that their levels are occupied up to their Fermi momentum $p_\chi$.
Their number density and pressure are
\beq n_\chi=  g\int\frac{d^3p}{(2\pi)^3}  =\frac{g p_\chi^3}{6\pi^2},
\qquad
\rho_\chi = g\int\frac{d^3p}{(2\pi)^3} E  ,\qquad
\wp_\chi = g\int\frac{d^3p}{(2\pi)^3}\frac{p^2}{3E}. \eeq
Thermal equilibrium of $n \leftrightarrow \chi\chi\chi$ relates
the $\chi$ chemical potential $\mu_\chi =\sqrt{M^2 + p_\chi^2}$
to the chemical potential of neutrons as in eq.\eq{muchi}, i.e.\
$\mu_\chi = \mu_n /3$.\footnote{While $n \to \chi\gamma$ gives a larger $\mu_\chi=\mu_n$.
Indeed a reaction $i + j \leftrightarrow i' + j'$ fast enough to be in thermal equilibrium
implies the relation $\mu_i + \mu_j =\mu_{i'}+\mu_{j'}$
among the  chemical potentials, independently of particle-physics details.
We recall the thermodynamic
relations at $T=0$ used later.
Particles are added with energy $E = \mu$ at the Fermi sphere, so $d\rho = \mu \, dn$.
The pressure $\wp$ is then given by 
\beq \label{eq:thermo}
\wp = -\left.\frac{ \partial U}{\partial V}\right|_T = n^2 \frac{\partial}{\partial n}\frac{\rho}{n}
=n \mu - \rho\eeq having used
$U = N\med{E}$ and $N=nV$.}

\begin{figure}[t]
$$\includegraphics[width=0.46\textwidth]{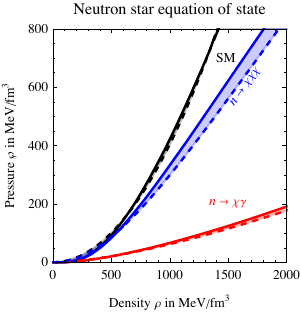}\qquad
\includegraphics[width=0.45\textwidth]{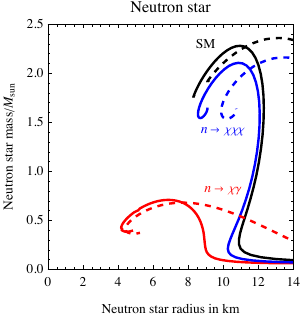}$$
 \caption{\label{fig:NS} {\bfseries Left:} Equations of state for neutron stars in the SM
 (black curves, considering two different computations of nuclear repulsion:
the continuous curve is  BSk24 from~\cite{1903.04981},
the dotted curve is  $\wp_n = \kappa_0 \rho_n^2$);
adding $n \leftrightarrow \chi\chi\chi$ (blue curves);
adding $n \leftrightarrow \chi\gamma$  (red curves).
 {\bfseries Right:} the corresponding relation between the radius and mass of neutron stars.
This shows that the observed neutron stars with mass around two solar masses are compatible with
$n \leftrightarrow \chi\chi\chi$, but not with $n \leftrightarrow \chi\gamma$.
The solutions below the peaks at smaller radii are unstable.}
\end{figure}

Concerning the SM particles (mostly neutrons), their equation
of state is significantly affected by nuclear repulsion, and
different approximations give somehow different results~\cite{nucl-th/0309041,1603.02698,
1707.04966,1903.04981,1904.04233,2012.03599}.
We adopt the following computations:
\begin{enumerate}
\item A first, rough but simple,
equation of state that includes nuclear effects is $\wp_n = \kappa_0 \rho_n^2$
with $\kappa_0 \approx 52/\GeV^4$~\cite{nucl-th/0309041}. 
The thermodynamic relation
$dn_n/n_n=d\rho_n/(\rho_n+\wp_n)$ of eq.\eq{thermo}
determines $n_n$ up to a constant,
fixed imposing $\mu_n = d\rho_n/dn_n \to m_n$ as $n_n\to 0$.
The result is
\beq \rho_n= \frac{m_n n_n}{1- m_n n_n \kappa_0},\qquad
\mu_n=\frac{m_n}{(1- m_n n \kappa_0)^2},\qquad
\wp_n=\kappa_0 \rho_n^2.\eeq
This equation of state is shown by the black dashed curve in fig.\fig{NS}a.

\item The possibly precise BSk24 equation of state from~\cite{1903.04981}, shown by the black curve in fig.\fig{NS}a,
agrees well with neutron star observations.
We use its numerical form, and present a rough analytic
Taylor approximation:  $\rho_n = a_1 n_n + a_2 n_n^2 + a_3 n_n^3$ with
$a_1 \approx m_n$, $a_2 \approx 16/\GeV^2$ and $a_3 \approx 4800/\GeV^5$.
Then thermodynamic relations imply
$\mu_n = m_n + 2 a_2 n_n + 3a_3 n_n^2$ and $\wp_n = a_2 n_n^2 + 2 a_3 n_n^3$.

\end{enumerate}
The red curves in fig.\fig{NS}a show how the equations of state computed in the SM
get significantly softened if $n \leftrightarrow \chi\gamma$ is in thermal equilibrium
(i.e.\ $\mu_n=\mu_\chi$) with $M=m_n$.
The blue curves  in fig.\fig{NS}a show how 
thermal equilibrium of $n \leftrightarrow \chi\chi\chi$
(i.e.\ $\mu_n=3\mu_\chi$) with $M=m_n/3$ leads to a milder modification of
the equation of state, that remains almost as hard as in the SM.

\begin{figure}[t]
$$\includegraphics[width=0.45\textwidth]{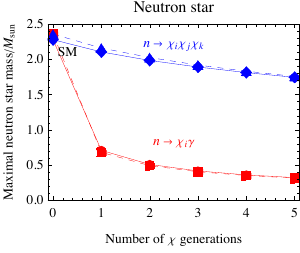}$$
 \caption{\label{fig:NS2} Maximal neutron star mass allowing for multiple $\chi$ particles
 with mass $M \approx m_n/3$ such that $n \to \chi\chi\chi$ is in thermal equilibrium (blue) 
 or mass $M \approx m_n$ such that $n \to \gamma\chi$ is in thermal equilibrium (red).  
The continuous and dashed curves again refer to two different computations of the
neutron equation of state.}
\end{figure}

As a result, fig.\fig{NS}b shows that the 
relation between the neutron star mass ${\cal M}$ and radius $R$
in the presence of $n \leftrightarrow \chi\chi\chi$
remains close enough to the SM limit
(in particular allowing for observed neutron stars with  
mass ${\cal M} \approx 2 M_{\rm sun}$),
unlike what happens if $n \leftrightarrow \chi\gamma$ is in thermal equilibrium.
In particular, while $n \leftrightarrow \chi\gamma$ reduces
the maximal mass of neutron stars in contradiction with data,
$n \leftrightarrow \chi\chi\chi$ leads to a 
mild reduction comparable to current SM uncertainties.
The neutron star radius is reduced in a similarly mild way compatible with data.
Different SM computations lead to maximal neutron star masses 
between $1.8M_{\rm sun}$ and  $2.6M_{\rm sun}$
and minimal radii between 10 km and 14 km~\cite{1603.02698}
in apparent agreement with data.
Thereby, if the SM can account for observed neutron stars, its $n\to\chi\chi\chi$ extension can too.
More precise future computations and observations
might be able to test the mild difference.

\smallskip

So far we considered $N=1$ generation of $\chi$ particles.
Fig.\fig{NS2} shows the maximal neutron star mass
as function of the number $N$ of $\chi$ generations, assuming for simplicity they all have the same mass.
We see that  $n\leftrightarrow \chi\chi\chi$ leads to such a small reduction that also $N>1$ is allowed.
The SM corresponds to $N=0$.

\section{Possible theories}\label{TH}
Having established that $n \leftrightarrow \chi\chi\chi$ is compatible with neutron star bounds,
we study its possible theory.
Since conserved $\chi$ number is needed,
$\chi$ must be a complex fermion(s) described by Dirac spinor(s) 
$\Psi = (\chi_L,\bar\chi_R)$ containing
two Weyl spinors $\chi_L$ and $\chi_R$.
The $n \to \chi\chi\chi$ decay arises from 4-fermion effective operators of the form
\beq \label{eq:n3chi}
\Lag_{\rm eff}=\Lag_{\rm SM}+
\bar \Psi (i\slashed{\partial}-M)\Psi+ \frac{(\bar\Psi^c \Gamma \Psi)(\bar n \Gamma \Psi)+ \hbox{h.c.}}{3!\Lambda^2_{\chi n}}
\eeq 
at nucleon level. 
This unusual operator involves the charge-conjugated field  $\Psi^c = C \bar\Psi^T$
such that the neutron decays into  three $\chi$ particles and no $\bar\chi$ anti-particles.
In the relevant non-relativistic limit $M\circa{<} m_n/3$ the decay rate is given by 
\beq \Gamma_{n\to\chi\chi\chi} 
 \simeq 
\frac{|\mathscr{A}|^2_{\rm nr} m_n}{(2\pi)^3 16 }
\left(1 -  \frac{3M}{m_n}\right)^2
= \frac{m_n^5}{ 27 \pi^3 \Lambda_{\chi n}^4}
\left\{\begin{array}{ll}
g_L^2 g_R^2 (1-3M/m_n)^3/16 & \hbox{if $\Gamma=g_L P_L + g_R P_R$}\\
g_A^2 (1-3M/m_n)^3 & \hbox{if $\Gamma=\gamma_\mu(g_V+ g_A\gamma_5)$}\\
{\cal O}(1-3M/m_n)^2 & \hbox{if $N>1$}\\
\end{array}
\right. 
\eeq
where $\mathscr{A}$  is the $n \leftrightarrow \chi\chi\chi$  decay amplitude. 
A minimal non-relativistic suppression 
$|\mathscr{A}|^2_{\rm nr} \equiv (m_n/\Lambda_n)^4 \epsilon$ with $\epsilon \sim 1- 3M/m_n$
necessarily arises if the decay is into $N=1$ $\chi$ generation.
This can be seen from non-relativistic quantum mechanics:
if the $\chi$ momenta vanish, the Pauli principle demands
the fully anti-symmetric product of three $\chi$ spinors
(heavy-quark effective theory can be adapted to obtain a systematic expansion~\cite{hep-ph/9306320}).
On the other hand, decays involving $N>1$ generations of $\chi$ particles allow for $\epsilon\sim 1$.
We thereby estimate
\beq \label{eq:Gnccc}
\Gamma_{n\to\chi\chi\chi}
\approx \epsilon \frac{m_n^5(1-3M/m_n)^2}{128\pi^3\Lambda_{\chi n}^4} \sim \epsilon\,  \Delta\Gamma
 \left(\frac{100\TeV}{\Lambda_{\chi n}}\right)^{4}\left(\frac{m_n - 3 M}{E_{\rm Be}}\right)^2.
\eeq
A lower $\Lambda_{\chi n} \approx 30\TeV$ is needed if $\epsilon \sim 1 - 3 M/m_n$.

\medskip

The rate of the related visible decay channel with an extra $\gamma$  
produced from the neutron magnetic moment interaction is estimated as
\beq \Gamma_{n\to\chi\chi\chi\gamma}  \approx \frac{\alpha E_\gamma^2}{4\pi m_n^2}\
\Gamma_{n\to\chi\chi\chi} \sim 10^{-9} \Gamma_{n\to\chi\chi\chi} ,\qquad
E_\gamma \circa{<} m_n - 3M
\eeq
and is safely below the experimental bounds~\cite{1802.01595}.

\medskip

In the present model $\chi$ can be DM while preserving hydrogen stability:
hydrogen decay is kinematically open only in the part of the
parameter space with lower $M$, as shown in fig.\fig{kinematics3chi}.
In such a case, the hydrogen decay rate can be estimated taking into account weak interactions of neutrons, 
\beq \Gamma_{{\rm H}  \to \chi\chi\chi\nu_e} \approx |\psi(0)|^2 \GF^2 E_\nu^2 \frac{ \Gamma_{n\to \chi\chi\chi} }{m_n}
\approx  \frac{1}{10^{30}\,{\rm yr}}\eeq
where $|\psi(0)|^2 \approx \alpha^3 m_e^3/\pi$ is the inverse atomic volume,
and $E_\nu \circa{<} m_p + m_e -3 M$.
Since this dominant hydrogen decay mode is fully invisible,
the experimental bound on its rate is comparable to the inverse universe age and largely satisfied.
A mildly stronger bound $\Gamma_{{\rm H}  \to \chi\chi\chi\nu_e}\circa{<}1/(10^{14}\,{\rm yr})$
arises considering electron $ep$ capture in the sun~\cite{1812.11089}.
Adding an extra photon gives a sub-dominant but visible hydrogen decay mode. Its rate 
\beq \Gamma_{{\rm H}  \to \chi\chi\chi\nu_e \gamma} \approx \frac{\alpha E_\gamma^2}{4\pi m_n^2} \Gamma_{{\rm H}  \to \chi\chi\chi\nu_e}
\approx \frac{1}{10^{39}\,{\rm yr}}\eeq
is compatible with the bound from {\sc Borexino},
$ \Gamma_{{\rm H}  \to \chi\chi\chi\nu_e \gamma}\circa{<} 1/(10^{28}\,{\rm yr})$~\cite{1509.01223,2003.02270}.

\bigskip

Coming back to theory,
the nucleon-level operator of eq.\eq{n3chi}
can arise from 6-fermion quark-level operators invariant under  SM
gauge interactions, such as 
\beq \label{eq:dim9}
\chi\chi\chi d_R d_R u_R /3!\Lambda^5_{q\chi}, \qquad 
 \chi\chi\chi d_R \bar Q \bar Q  /3!\Lambda^5_{q\chi}\eeq
where $Q=(u,d)_L$, $d_R$, $u_R$ are the SM quarks in Weyl notation and we omitted
$\chi$ chiralities, $\chi \sim \chi_L \sim \bar\chi_R$ and
Lorentz indices that can be contracted in multiple ways. 
Lattice computations of the nuclear matrix element
$\langle 0 | (u d)_{L,R} d_{L,R}| n\rangle = \beta_{L,R} n$ find
$|\beta_{L,R}| \approx 0.014\GeV^3 \sim  \Lambda_{\rm QCD}^3$~\cite{1705.01338},
and chiral perturbation theory allows to compute interactions with extra pions or photons~\cite{Claudson:1981gh}.

\smallskip

In view of their large dimension 9, the operators in eq.\eq{dim9} 
are strongly suppressed by $\Lambda_{q\chi}$.
The neutron-level operator of eq.\eq{n3chi} is obtained with coefficient
$1/\Lambda_{\chi n}^2 =\beta_{L,R}/\Lambda_{\chi q}^5$,
so that  the neutron decay anomaly is reproduced for low $\Lambda_{\chi q} \sim 30\GeV$.
Since $\Lambda_{\chi q} \sim {30}\GeV$ is below the weak scale,
a mediator below the weak scale is needed.
This can be achieved, compatibly with collider bounds, 
adding for example
a neutral fermion $n'$, singlet under all SM gauge interactions,
coupled via dimension-6 operators as 
\beq \label{eq:n'}\bar n'\chi\chi\chi/\Lambda_{\chi n'}^2 + \bar n' ddu/\Lambda_{nn'}^2.\eeq
Baryon number is conserved assigning $B_{n'}=1$  to $n'$.
In turn, the 4-fermion operators in eq.\eq{n'} can be mediated by renormalizable couplings of
bosons, such as
a $W_R$ or a lepto-quark, and by their dark-sector analogous.
One gets $\Lambda^5_n = M_{n'} \Lambda_{\chi n'}^2\Lambda_{nn'}^2$, so that
the 4-fermion operator $\bar n' ddu$ involving SM particles can be suppressed by a
$\Lambda_{nn'}$ above the weak scale, up to a few $\TeV$,
while keeping $M_{n'}>1.5 m_n$ in order to avoid conflicting with neutron star bounds~\cite{1802.08282},
and keeping 
the dark interaction $\Lambda_{\chi n'}$ above the QCD scale in order to avoid
conflicting with bounds on DM self-interactions,
$\sigma/M \circa{<} 10^4/\GeV^3$~\cite{astro-ph/0608407,1308.3419}.

\medskip

\section{Dark Matter cosmological abundance}\label{DMOmega}
To match the observed cosmological DM abundance, the $\chi$ particles must have abundance
$Y \equiv n/s = 0.44\eV/M = 1.4~10^{-9}$,
where $n$ is the $\chi + \bar\chi$ number density and $s = 2\pi^3 d_{\rm SM} T^3/45$ 
is the entropy density of the thermal bath
with $d_{\rm SM}(T)$ degrees of freedom at temperature $T$.

The neutron decay anomaly is reproduced for $\Lambda_{\chi n} \gg v$,
so the interaction rates of $\chi$ particles at $T \sim m_n$ are below 
electroweak rates and thereby below the Hubble rate:
thermal freeze-out is not possible.\footnote{Ignoring the neutron decay anomaly,
one could consider the regime $m_n/3< M< m_n$ where neutron decay is kinematically closed, 
allowing for larger DM couplings.}
For the same reason, no equilibration of asymmetries happens at $T \circa{<} m_n$.

We thereby consider freeze-in, assuming vanishing initial $\chi$ abundance
and computing the multiple contributions it receives from particle-physics processes.
We avoid entering into model-dependent mediator issues, and explore the effect of the interaction with neutrons in eq.\eq{n3chi}, for example by assuming that the reheating temperature is below the mediator masses.
At leading order, neutrons and anti-neutrons that decay at $T \sim m_n$ close to thermal equilibrium\footnote{We can neglect the later out-of-equilibrium decays at BBN of neutrons that remained thanks to the baryon asymmetry:
as the neutron BR into DM is about $1\%$, their freeze-in contribution to the DM abundance is too small.}
contribute to the DM abundance  as
\beq Y = \frac{405\sqrt{5} M_{\rm Pl} }{4\pi^{9/2} d_{\rm SM}^{3/2} m_n^2} 3 \Gamma_{n\to\chi\chi\chi} R
\approx 10^{-13}
\eeq
where $R \sim e^{-\Lambda_{\rm QCD}/m_n}$ corrects 
the freeze-in formula taking into
account  that neutrons only form after the QCD phase transition at $T \circa{<} \Lambda_{\rm QCD}$.
This $Y$ is about $10^4$ smaller than the cosmological DM abundance.
Three different effects provide the needed enhancement.

\begin{enumerate}
\item First, thermal scatterings such as $\pi^0 n \leftrightarrow \chi\chi\chi$ and
$\gamma n \leftrightarrow \chi\chi\chi$ arise
at higher orders in the QCD or QED coupling $g$, and
avoid the non-relativistic suppression of $n \leftrightarrow \chi\chi\chi$.
This suppression was computed in eq.\eq{Gnccc}: two powers of $1-3M/m_n \approx \hbox{few}\times 10^{-3}$ 
arise from the phase space,
and  possibly one extra power arises from the squared amplitude.
Thereby, at the relevant temperature $T \sim m_n$, the scattering 
rates are enhanced by a factor of order $g^2 (T/E_{\rm Be})^{2-3} \sim 10^{4-6}$
compared to the $\Gamma_{n\to\chi\chi\chi}$ rate.
Indeed these finite-temperature scatterings can be partially accounted by a 
$n\leftrightarrow\chi\chi\chi$ rate enhanced by taking into account the thermal contribution to the 
neutron squared  mass, $m_n^2 + {\cal O}(g^2 T^2)$.

\item
Second, $n^* \to \chi\chi\chi$ decays of the excited neutron $n^*$ with mass $m_{n^*}\approx 1.44\GeV$
and of other QCD resonances similarly avoid the non-relativistic suppression of $n\to \chi\chi\chi$.
Dedicated lattice computations are needed to confirm that these resonances have  matrix
elements similar to those of the neutron.
We expect that their effect is numerically similar to thermal scatterings at point 1.
\item
Third, the processes $\bar \chi n \leftrightarrow\chi\chi$ and $ \chi \bar n \leftrightarrow \bar \chi\bar\chi$
have similar
phase-space unsuppressed rates, and increase the number of 
DM particles that have already been produced, adding $\bar\chi$ anti-particles.
Their effect is numerically comparable to those at previous points.
\end{enumerate}
A precise computation is not possible since non-perturbative QCD effects determine the rates at point 2.
Furthermore, larger but model-dependent contributions to freeze-in can arise at higher temperature,
if the reheating temperature is higher than $\Lambda_{\rm QCD}$.

\section{Dark Matter signals}\label{DMsig}
The $\bar \chi n \leftrightarrow\chi\chi$ process at point 3 leads to unusual DM direct detection signals:
it is kinematically open today and the non-relativistic $\bar\chi$ and $n$ in its initial state
produce relativistic $\chi$ with energy $E = 2 m_n/3$.
The $\bar \chi n \leftrightarrow\chi\chi$ cross section avoids the non-relativistic suppression of $n\to\chi\chi\chi$
and is thereby estimated as 
\beq \label{eq:DDsigma}
\sigma_{\bar \chi n \to\chi\chi } \approx \frac{m_n^2}{4\pi \Lambda_{\chi n}^4} \sim 10^{-46}\cm^2 \, 
\left(\frac{20\TeV}{\Lambda_n}\right)^4.\eeq
This is below current bounds from direct detection experiments
for the value of $\Lambda_{\chi n}$ motivated by the neutron decay anomaly.
No nuclear enhancement arises, since
the energies involved in $\bar \chi n \leftrightarrow\chi\chi$ are higher than 
the tens of keV produced by usual DM-induced nuclear recoils.
For the same reason, bigger experiments dedicated to more energetic signals produced by 
neutrinos and proton decay have better sensitivity.
Following their practice, we convert eq.\eq{DDsigma} into
the event rate per neutron 
as
\beq \label{eq:taundisap}
\tau= 
\frac{\rho_{\odot}}{2M} v \sigma_{\bar \chi n \to\chi\chi} =
\frac{1}{2.5~10^{31}\,{\rm yr}}
\frac{\sigma_{\bar \chi n \to\chi\chi}}{10^{-46}\cm^2} 
\frac{\rho_{\odot}}{0.4\,{\GeV}/{\cm^3}} 
\frac{v}{200\, {\rm km}/{\rm s}}
 \frac{m_n/3}{M}
\eeq
having assumed an equal number of $\chi$ and $\bar \chi$.
We thereby have the following signals:
\begin{enumerate}
\item {\em DM scatterings that lead to the disappearance of a neutron into invisible DM}.
These scatterings effectively make ordinary matter radioactive with life-time given by eq.\eq{taundisap},
 since a neutron that disappears within a nucleus leaves a hole, triggering nuclear 
de-excitations and decays. 
For example DM scatterings convert water $^{16}{\rm O}_8$ into $^{15}{\rm O}_8$,
that de-excites emitting $\gamma$ rays with tens of MeV, and
decays a few minutes later  into $^{15}{\rm N}_7 \, e^+\nu_e$ emitting a $\sim \MeV$  positron. 
Similar processes happen with $^{12}{\rm C}_{6}$.
The experimental bound on DM-induced neutron disappearance
is approximatively obtained recasting the bounds on
invisible neutron decays from experiments done with $^{16}{\rm O}_8$ or $^{12}{\rm C}_{6}$:
\beq \tau(n\to \hbox{invisible}) >\left\{\begin{array}{ll}
4.9~10^{26} \,{\rm yr}& \hbox{from {\sc Kamiokande}~\cite{Kamiokande:1993ivj}}\\
2.5~10^{29}\,{\rm yr} & \hbox{from SNO~\cite{1812.05552}}\\
5.8~10^{29}\,{\rm yr} & \hbox{from {\sc KamLand}~\cite{KamLAND:2005pen}}
\end{array}\right.  .\eeq
The stronger  {\sc SuperKamiokande}~\cite{1508.05530} bound on $n p \to e^+ \nu$ 
cannot be recasted into a bound on neutron disappearance,
as the experimental search  triggered on more energetic positrons with $p>100 \MeV$.


\item {\em DM scatterings that lead to the disappearance of a neutron into invisible DM 
plus visible SM particles}. Processes like
$\bar\chi n \to \chi \chi \gamma$ and
$\bar\chi n \to \chi \chi \pi^0 \to \chi \chi \gamma\gamma $  
arise at higher orders and thereby have cross sections (and consequently effective neutron decay rates)
smaller by $\alpha/4\pi \sim 10^{-2-3}$ compared to the fully invisible neutron disappearance.
The extra photons have energy around $100\MeV$.
While a precise recasting needs computing the photon spectra,
these processes are loosely similar to $n\to \nu \gamma$, subject to the bound
\beq \tau(n\to \nu \gamma) >5.5~10^{32}\,{\rm yr}\qquad
\hbox{from {\sc SuperKamiokande}~\cite{1508.05530}}\eeq
that is thereby satisfied.
\end{enumerate}
Concerning indirect detection signals, extra interactions not needed by the model
might lead to $\chi\bar \chi$ annihilations into pairs of SM particles.

\section{Conclusions}\label{concl}
Past literature showed that
the interpretation of the neutron decay anomaly in terms
of $n\to \chi\gamma$ decays into new free invisible particles $\chi$ with baryon number $B_\chi=1$
implies their thermalization inside neutron stars,
and that the modified equation of state is in contradiction with observed neutron stars.

\medskip

We found that the neutron star issue is more general, affecting all possible 
models listed in table~\ref{tab:BLmodels} where we 
considered generic $\chi$ charges under conserved baryon and lepton numbers, $B_\chi$ and $L_\chi$.
Independently of particle-physics details,
the $\chi$ chemical potential in thermal equilibrium is $\mu_\chi = B_\chi \mu_n$.
Neutron star bounds are found to be compatible with the 
minimal possibility that $\chi$ carries the smallest baryon number $B_\chi=1/3$,
so that the neutron decay anomaly can be accounted by $n \to\chi\chi\chi$ decays.
The resulting modification of neutron stars, shown in  fig.\fig{NS}, is 
comparable with SM uncertainties.
Fig.\fig{NS2} shows that neutron stars get modified in a specific and mild enough way
even in the presence of multiple generations of $\chi$ particles,
thereby providing a signal for improved future observations and computations.

\smallskip

In section~\ref{TH} we presented possible theories for $n\to\chi\chi\chi$,
showing that associated processes are compatible with bounds from colliders,
on  $n\to\chi\chi\chi\gamma$  decays,
and on hydrogen decays into $\chi\chi\chi \nu_e$ or $\chi\chi\chi \nu_e \gamma$
(kinematically open only in a part of parameter space shown in fig.\fig{kinematics3chi}).

\smallskip

We next found that  $\chi$ can be Dark Matter.
In section~\ref{DMOmega} we showed that the cosmological DM abundance
can be possibly matched by freeze-in production of $\chi$
 after the QCD phase transition at $T \sim m_n$.
Three processes contribute: 
$\pi^0 n \leftrightarrow \chi\chi\chi$ and $\gamma n \leftrightarrow \chi\chi\chi$ thermal scatterings;
decays of excited neutrons;
$ \chi \bar n \leftrightarrow \bar \chi\bar\chi$ and $\bar\chi n \leftrightarrow \chi\chi$ scatterings.
In section~\ref{DMsig} we found that 
$\bar\chi n \leftrightarrow \chi\chi$ provide unusual DM signals with interesting rates below present bounds.
Such DM scatterings lead to neutron disappearance,
that effectively makes matter (such as water, carbon, etc) radioactive,
giving signatures similar to those of neutron invisible decays.
Predictions are compatible with bounds from DM detection experiments and
from neutrino experiments such as SNO and {\sc KamLand}.
Furthermore, the higher order related processes with  extra visible photons,
$\bar\chi n \to \chi\chi \gamma$
and $\bar\chi n \to \chi\chi \pi^0\to\chi\chi \gamma \gamma$,
are compatible with {\sc Super-Kamiokande} bounds.

\small

\paragraph{Acknowledgements}
This work was supported by MIUR under PRIN 2017FNJFMW
and by the ERC grant 669668 NEO-NAT.
We thank Zurab Berezhiani, Raghuveer Garani,
Benjamin Grinstein, Masayuki Nakahata, Michele Redi
 and Daniele Teresi for discussions.

\small\appendix


\footnotesize

\begin{thebibliography}{nnn}\bibitem{nucl-ex/0408009}
\article[nucl-ex/0408009]{A. Serebrov et al.}{Phys.Lett.B}{605}{72}{2005}
{\href{https://doi.org/10.1016/j.physletb.2004.11.013}{Measurement of the neutron lifetime using a gravitational trap and a low-temperature Fomblin coating}}.


\bibitem{Pichlmaier:2010zz}
\article{A. Pichlmaier, V. Varlamov, K. Schreckenbach, P. Geltenbort}{Phys.Lett.B}{693}{221}{2010}
{\href{https://doi.org/10.1016/j.physletb.2010.08.032}{Neutron lifetime measurement with the UCN trap-in-trap MAMBO II}}.


\bibitem{Steyerl:2012zz}
\article{A. Steyerl, J.M. Pendlebury, C. Kaufman, S.S. Malik, A.M. Desai}{Phys.Rev.C}{85}{065503}{2012}
{\href{https://doi.org/10.1103/PhysRevC.85.065503}{Quasielastic scattering in the interaction of ultracold neutrons with a liquid wall and application in a reanalysis of the Mambo I neutron-lifetime experiment}}.


\bibitem{1412.7434}
\article[1412.7434]{V.F. Ezhov et al.}{JETP Lett.}{107}{671}{2018}
{\href{https://doi.org/10.1134/S0021364018110024}{Measurement of the neutron lifetime with ultra-cold neutrons stored in a magneto-gravitational trap}}.


\bibitem{Arzumanov:2015tea}
\article{S. Arzumanov, L. Bondarenko, S. Chernyavsky, P. Geltenbort, V. Morozov, V.V. Nesvizhevsky, Y. Panin, A. Strepetov}{Phys.Lett.B}{745}{79}{2015}
{\href{https://doi.org/10.1016/j.physletb.2015.04.021}{A measurement of the neutron lifetime using the method of storage of ultracold neutrons and detection of inelastically up-scattered neutrons}}.


\bibitem{1707.01817}
\article[1707.01817]{R.W. Pattie Jr. et al.}{Science}{360}{627}{2018}
{\href{https://doi.org/10.1126/science.aan8895}{Measurement of the neutron lifetime using a magneto-gravitational trap and in situ detection}}.


\bibitem{1712.05663}
\article[1712.05663]{A. P. Serebrov et al.}{Phys.Rev.C}{97}{055503}{2018}
{\href{https://doi.org/10.1103/PhysRevC.97.055503}{Neutron lifetime measurements with a large gravitational trap for ultracold neutrons}}.


\bibitem{2106.10375}
\article[2106.10375]{{\sc UCN$\tau$} Collaboration}{Phys.Rev.Lett.}{127}{162501}{2021}
{\href{https://doi.org/10.1103/PhysRevLett.127.162501}{Improved Neutron Lifetime Measurement with UCN$\tau$}}.


\bibitem{Byrne:1996zz}
\article{J. Byrne, P.G. Dawber}{Europhys.Lett.}{33}{187}{1996}
{\href{https://doi.org/10.1209/epl/i1996-00319-x}{A Revised Value for the Neutron Lifetime Measured Using a Penning Trap}}.


\bibitem{nucl-ex/0411041}
\article[nucl-ex/0411041]{J.S. Nico et al.}{Phys.Rev.C}{71}{055502}{2005}
{\href{https://doi.org/10.1103/PhysRevC.71.055502}{Measurement of the neutron lifetime by counting trapped protons in a cold neutron beam}}.


\bibitem{1309.2623}
\article[1309.2623]{A.T. Yue, M.S. Dewey, D.M. Gilliam, G.L. Greene, A.B. Laptev, J.S. Nico, W.M. Snow, F.E. Wietfeldt}{Phys.Rev.Lett.}{111}{222501}{2013}
{\href{https://doi.org/10.1103/PhysRevLett.111.222501}{Improved Determination of the Neutron Lifetime}}.


\bibitem{1801.01124}
\article[1801.01124]{B. Fornal, B. Grinstein}{Phys.Rev.Lett.}{120}{191801}{2018}
{\href{https://doi.org/10.1103/PhysRevLett.120.191801}{Dark Matter Interpretation of the Neutron Decay Anomaly}}.


\bibitem{2007.13931}
\article[2007.13931]{B. Fornal, B. Grinstein}{Mod.Phys.Lett.A}{35}{2030019}{2020}
{\href{https://doi.org/10.1142/S0217732320300190}{Neutron's dark secret}}.


\bibitem{Zurabtalk}
Z. Berezhiani, \href{https://www.int.washington.edu/talks/WorkShops/int_17_69W/People/Berezhiani_Z/Berezhiani3.pdf}{talk at INT, Seattle, 22-27 Oct.\ 2017}.


\bibitem{1004.2981}
\article[1004.2981]{A.P. Serebrov, O.M. Zherebtsov}{Astron.Lett.}{37}{181}{2011}
{\href{https://doi.org/10.1134/S1063773711010075}{Trap with ultracold neutrons as a detector of dark matter particles with long-range forces}}.


\bibitem{2008.06061}
\heparticle[2008.06061]{S. Rajendran, H. Ramani}{A Composite Solution to the Neutron Bottle Anomaly}.


\bibitem{1802.01595}
\article[1802.01595]{Z. Tang et al.}{Phys.Rev.Lett.}{121}{022505}{2018}
{\href{https://doi.org/10.1103/PhysRevLett.121.022505}{Search for the Neutron Decay n$\rightarrow$ X+$\gamma$ where X is a dark matter particle}}.


\bibitem{1509.01223}
\article[1509.01223]{{\sc Borexino} Collaboration}{Phys.Rev.Lett.}{115}{231802}{2015}
{\href{https://doi.org/10.1103/PhysRevLett.115.231802}{A test of electric charge conservation with Borexino}}.


\bibitem{Oppenheimer:1939ne}
\article{J.R. Oppenheimer, G.M. Volkoff}{Phys.Rev.}{55}{374}{1939}
{\href{https://doi.org/10.1103/PhysRev.55.374}{On massive neutron cores}}.


\bibitem{1802.01804}
\article[1802.01804]{A. Czarnecki, W.J. Marciano, A. Sirlin}{Phys.Rev.Lett.}{120}{202002}{2018}
{\href{https://doi.org/10.1103/PhysRevLett.120.202002}{Neutron Lifetime and Axial Coupling Connection}}.


\bibitem{Berezhiani:2018eds}
\article[1807.07906]{Z. Berezhiani}{Eur.Phys.J.C}{79}{484}{2019}
{\href{https://doi.org/10.1140/epjc/s10052-019-6995-x}{Neutron lifetime puzzle and neutron/mirror neutron oscillation}}.


\bibitem{Berezhiani:2020zck}
\article[2012.15233]{Z. Berezhiani, R. Biondi, M. Mannarelli, F. Tonelli}{Eur.Phys.J.C}{81}{1036}{2021}
{\href{https://doi.org/10.1140/epjc/s10052-021-09806-1}{Neutron-mirror neutron mixing and neutron stars}}.


\bibitem{Berezhiani:2000gw}
\article[hep-ph/0008105]{Z. Berezhiani, D. Comelli, F.L. Villante}{Phys.Lett.B}{503}{362}{2001}
{\href{https://doi.org/10.1016/S0370-2693(01)00217-9}{The Early mirror universe: Inflation, baryogenesis, nucleosynthesis and dark matter}}.


\bibitem{Lee:1956qn}
\article{T.D. Lee, C.-N. Yang}{Phys.Rev.}{104}{254}{1956}
{\href{https://doi.org/10.1103/PhysRev.104.254}{Question of Parity Conservation in Weak Interactions}}.


\bibitem{Kobzarev:1966qya}
\article{I.Y. Kobzarev, L.B. Okun, I.Y. Pomeranchuk}{Sov.J.Nucl.Phys.}{3}{837}{1966}
{On the possibility of experimental observation of mirror particles}.


\bibitem{Blinnikov:1983gh}
\article{S.I. Blinnikov, M. Khlopov}{Sov.Astron.}{27}{371}{1983}
{Possible astronomical effects of mirror particles}.


\bibitem{Kolb:1985bf}
\article{E.W. Kolb, D. Seckel, M.S. Turner}{Nature}{314}{415}{1985}
{\href{https://doi.org/10.1038/314415a0}{The Shadow World}}.


\bibitem{Foot:1991py}
\article{R. Foot, H. Lew, R.R. Volkas}{Mod.Phys.Lett.A}{7}{2567}{1992}
{\href{https://doi.org/10.1142/S0217732392004031}{Possible consequences of parity conservation}}.


\bibitem{Hodges:1993yb}
\article{H.M. Hodges}{Phys.Rev.D}{47}{456}{1993}
{\href{https://doi.org/10.1103/PhysRevD.47.456}{Mirror baryons as the dark matter}}.


\bibitem{hep-ph/9511221}
\article[hep-ph/9511221]{Z.G. Berezhiani, A.D. Dolgov, R.N. Mohapatra}{Phys.Lett.B}{375}{26}{1996}
{\href{https://doi.org/10.1016/0370-2693(96)00219-5}{Asymmetric inflationary reheating and the nature of mirror universe}}.


\bibitem{hep-ph/9602326}
\article[hep-ph/9602326]{Z.G. Berezhiani}{Acta Phys.Polon.B}{27}{1503}{1996}
{Astrophysical implications of the mirror world with broken mirror parity}.


\bibitem{1303.1727}
\article[1303.1727]{R. Foot}{Phys.Dark Univ.}{5-6}{236}{2014}
{\href{https://doi.org/10.1016/j.dark.2014.05.007}{A dark matter scaling relation from mirror dark matter}}.


\bibitem{1401.3965}
\article[1401.3965]{R. Foot}{Int.J.Mod.Phys.A}{29}{1430013}{2014}
{\href{https://doi.org/10.1142/S0217751X14300130}{Mirror dark matter: Cosmology, galaxy structure and direct detection}}.


\bibitem{hep-ph/0507031}
\article[hep-ph/0507031]{Z. Berezhiani, L. Bento}{Phys.Rev.Lett.}{96}{081801}{2006}
{\href{https://doi.org/10.1103/PhysRevLett.96.081801}{Neutron - mirror neutron oscillations: How fast might they be?}}.


\bibitem{1802.08244}
\article[1802.08244]{D. McKeen, A.E. Nelson, S. Reddy, D. Zhou}{Phys.Rev.Lett.}{121}{061802}{2018}
{\href{https://doi.org/10.1103/PhysRevLett.121.061802}{Neutron stars exclude light dark baryons}}.


\bibitem{1802.08282}
\article[1802.08282]{G. Baym, D.H. Beck, P. Geltenbort, J. Shelton}{Phys.Rev.Lett.}{121}{061801}{2018}
{\href{https://doi.org/10.1103/PhysRevLett.121.061801}{Testing dark decays of baryons in neutron stars}}.


\bibitem{1802.08427}
\article[1802.08427]{T.F. Motta, P.A.M. Guichon, A.W. Thomas}{J.Phys.G}{45}{05LT01}{2018}
{\href{https://doi.org/10.1088/1361-6471/aab689}{Implications of Neutron Star Properties for the Existence of Light Dark Matter}}.


\bibitem{1811.06546}
\article[1811.06546]{B. Grinstein, C. Kouvaris, N.G. Nielsen}{Phys.Rev.Lett.}{123}{091601}{2019}
{\href{https://doi.org/10.1103/PhysRevLett.123.091601}{Neutron Star Stability in Light of the Neutron Decay Anomaly}}.


\bibitem{0806.0876}
\article[0806.0876]{F. del Aguila, S. Bar-Shalom, A. Soni, J. Wudka}{Phys.Lett.B}{670}{399}{2009}
{\href{https://doi.org/10.1016/j.physletb.2008.11.031}{Heavy Majorana Neutrinos in the Effective Lagrangian Description: Application to Hadron Colliders}}.


\bibitem{1410.4193}
\article[1410.4193]{L. Lehman}{Phys.Rev.D}{90}{125023}{2014}
{\href{https://doi.org/10.1103/PhysRevD.90.125023}{Extending the Standard Model Effective Field Theory with the Complete Set of Dimension-7 Operators}}.


\bibitem{2005.00008}
\article[2005.00008]{H.-L. Li, Z. Ren, J. Shu, M.-L. Xiao, J.-H. Yu, Y.-H. Zheng}{Phys.Rev.D}{104}{015026}{2021}
{\href{https://doi.org/10.1103/PhysRevD.104.015026}{Complete set of dimension-eight operators in the standard model effective field theory}}.


\bibitem{2007.07899}
\article[2007.07899]{H.-L. Li, Z. Ren, M.-L. Xiao, J.-H. Yu, Y.-H. Zheng}{Phys.Rev.D}{104}{015025}{2021}
{\href{https://doi.org/10.1103/PhysRevD.104.015025}{Complete set of dimension-nine operators in the standard model effective field theory}}.


\bibitem{2007.08125}
\article[2007.08125]{Y. Liao, X.-D. Ma}{JHEP}{152}{}{2020}
{\href{https://doi.org/10.1007/JHEP11(2020)152}{An explicit construction of the dimension-9 operator basis in the standard model effective field theory}}.


\bibitem{astro-ph/0605724}
\article[astro-ph/0605724]{G. Narain, J. Schaffner-Bielich, I.N. Mishustin}{Phys.Rev.D}{74}{063003}{2006}
{\href{https://doi.org/10.1103/PhysRevD.74.063003}{Compact stars made of fermionic dark matter}}.


\bibitem{1903.04981}
\article[1903.04981]{J.M. Pearson, N. Chamel, A.Y. Potekhin, A.F. Fantina, C. Ducoin, A.K. Dutta, S. Goriely}{Mon.Not.Roy.Astron.Soc.}{481}{2994}{2018}
{\href{https://doi.org/10.1093/mnras/sty2413}{Unified equations of state for cold non-accreting neutron stars with Brussels–Montreal functionals – I. Role of symmetry energy}}.


\bibitem{nucl-th/0309041}
\heparticle[nucl-th/0309041]{R.R. Silbar, S. Reddy}{Neutron stars for undergraduates}.


\bibitem{1603.02698}
\article[1603.02698]{F. {\" O}zel, P. Freire}{Ann.Rev.Astron.Astrophys.}{54}{401}{2016}
{\href{https://doi.org/10.1146/annurev-astro-081915-023322}{Masses, Radii, and the Equation of State of Neutron Stars}}.


\bibitem{1707.04966}
\article[1707.04966]{G. Baym, T. Hatsuda, T. Kojo, P.D. Powell, Y. Song, T. Takatsuka}{Rept.Prog.Phys.}{81}{056902}{2018}
{\href{https://doi.org/10.1088/1361-6633/aaae14}{From hadrons to quarks in neutron stars: a review}}.


\bibitem{1904.04233}
\article[1904.04233]{M.M.N. Forbes, S. Bose, S. Reddy, D. Zhou, A. Mukherjee, S. De}{Phys.Rev.D}{100}{083010}{2019}
{\href{https://doi.org/10.1103/PhysRevD.100.083010}{Constraining the neutron-matter equation of state with gravitational waves}}.


\bibitem{2012.03599}
\article[2012.03599]{D. Logoteta, A. Perego, I. Bombaci}{Astron.Astrophys.}{646}{A55}{2021}
{\href{https://doi.org/10.1051/0004-6361/202039457}{Microscopic equation of state of hot nuclear matter for numerical relativity simulations}}.


\bibitem{hep-ph/9306320}
See e.g.\
\article[hep-ph/9306320]{M. Neubert}{Phys. Rept.}{245}{259}{1994}
{\href{https://doi.org/10.1016/0370-1573(94)90091-4}{Heavy quark symmetry}}.


\bibitem{1812.11089}
\article[1812.11089]{Z. Berezhiani}{LHEP}{2}{118}{2019}
{\href{https://doi.org/10.31526/lhep.1.2019.118}{Neutron lifetime and dark decay of the neutron and hydrogen}}.


\bibitem{2003.02270}
\heparticle[2003.02270]{D. McKeen, M. Pospelov}{How long does the hydrogen atom live?}.


\bibitem{1705.01338} 
\article[1705.01338]{Y. Aoki, T. Izubuchi, E. Shintani, A. Soni}{Phys.Rev.D}{96}{014506}{2017}
{\href{https://doi.org/10.1103/PhysRevD.96.014506}{Improved lattice computation of proton decay matrix elements}}.


\bibitem{Claudson:1981gh}
\article{M. Claudson, M.B. Wise, L.J. Hall}{Nucl.Phys.B}{195}{297}{1982}
{\href{https://doi.org/10.1016/0550-3213(82)90401-1}{Chiral Lagrangian for Deep Mine Physics}}.


\bibitem{astro-ph/0608407}
\article[astro-ph/0608407]{D. Clowe, M. Bradac, A.H. Gonzalez, M. Markevitch, S.W. Randall, C. Jones, D. Zaritsky}{Astrophys.J.Lett.}{648}{L109}{2006}
{\href{https://doi.org/10.1086/508162}{A direct empirical proof of the existence of dark matter}}.


\bibitem{1308.3419}
\article[1308.3419]{F. Kahlhoefer, K. Schmidt-Hoberg, M.T. Frandsen, S. Sarkar}{Mon.Not.Roy.Astron.Soc.}{437}{2865}{2014}
{\href{https://doi.org/10.1093/mnras/stt2097}{Colliding clusters and dark matter self-interactions}}.


\bibitem{Kamiokande:1993ivj}
\article{{\sc Kamiokande} Collaboration}{Phys.Lett.B}{311}{357}{1993}
{\href{https://doi.org/10.1016/0370-2693(93)90582-3}{Study of invisible nucleon decay, $n \to\nu\nu\bar\nu$, and a forbidden nuclear transition in the Kamiokande detector}}.


\bibitem{1812.05552}
\article[1812.05552]{{\sc SNO+} Collaboration}{Phys.Rev.D}{99}{032008}{2019}
{\href{https://doi.org/10.1103/PhysRevD.99.032008}{Search for invisible modes of nucleon decay in water with the SNO+ detector}}.


\bibitem{KamLAND:2005pen}
\article[hep-ex/0512059]{{\sc KamLAND} Collaboration}{Phys.Rev.Lett.}{96}{101802}{2006}
{\href{https://doi.org/10.1103/PhysRevLett.96.101802}{Search for the invisible decay of neutrons with KamLAND}}.


\bibitem{1508.05530}
\article[1508.05530]{{\sc Super-Kamiokande} Collaboration}{Phys.Rev.Lett.}{115}{121803}{2015}
{\href{https://doi.org/10.1103/PhysRevLett.115.121803}{Search for Nucleon and Dinucleon Decays with an Invisible Particle and a Charged Lepton in the Final State at the Super-Kamiokande Experiment}}.


\end{thebibliography}
\end{document}